\documentclass[twocolumn]{revtex4-1}
\usepackage{amsmath,bbm}
\usepackage{graphics}
\usepackage{epsfig}
\usepackage{epstopdf}

\newcommand{\ket}[2][]{{|#2\rangle_{#1}}}
\newcommand{\bra}[2][]{{}_{#1}\langle #2|}

\DeclareMathOperator{\sinc}{sinc}

\begin{document}
\title{Generation of Polarization Entangled Photon Pairs in a Planar Waveguide  }
\author{Divya Bharadwaj }
\email[]{divyabharadwaj4@gmail.com}
\altaffiliation{}
\author{K. Thyagarajan}

\affiliation{Department of Physics, IIT Delhi, New Delhi 110016, India}

\begin{abstract}
In this paper, we show that the polarization entangled photon pairs can be generated using spontaneous parametric down conversion in a dual periodically poled planar waveguide. The proposed configuration is shown to have efficiencies higher than in the case of bulk crystals with the possibility of ease of collection like in channel waveguides with the additional feature of wavelength tunability. We also show that the planar waveguide configuration permits the generation of hyper entanglement in polarization and path degree of freedom which is not possible in case of channel waveguide. The proposed design should find applications in quantum information processing using integrated quantum optics.
\end{abstract}
\maketitle

\section{Introduction}
Spontaneous parametric down conversion (SPDC) is one of the most extensively used processes for the generation of entangled photon pairs. Generation of entangled photon pairs have been extensively studied both in bulk crystals  \cite{1,2} and channel waveguides  \cite{3,4,5,6}. Entangled photon pairs generated through SPDC process in bulk nonlinear optic crystals suffer from a number of problems such as low efficiency, design complication, low interaction length, and are hard to implement for practical applications. In order to overcome these problems SPDC process has been implemented in nonlinear channel waveguides which have higher down conversion efficiencies in comparison to a bulk crystal \cite{3,4,5,6}.  At the same time, the flexibility in terms of variation in pump frequency or tunability in the generation of entangled pairs is not possible in channel waveguides.

In this aspect, planar waveguides which have confinement only along one direction can offer us better efficiency compared to bulk while at the same time providing us with a possibility of frequency tunability of the SPDC pairs by appropriately tuning the pump wavelength and ease of collection of the generated photon pairs. 

In this paper we present the design and analysis of the generation of polarization entangled photon pairs in planar waveguides and show the advantages of this vis a vis bulk and channel waveguide geometries. Numerical simulation for optimization of the entangled photon pair generation process and tunability of the generated photon pairs is carried out using planar waveguides in KTP.
\section{PRINCIPLE}
\label{Sec:PRINCIPLE}
\begin{figure}[t]
\centering
\includegraphics[width=8cm]{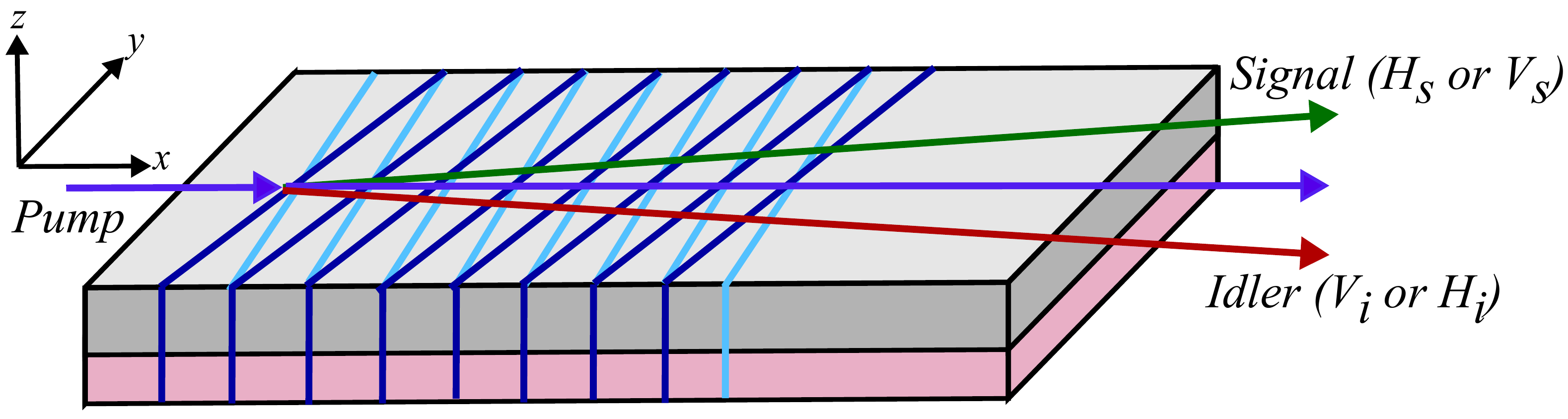}
\caption{Schematic of planar waveguide with the two slant periodically poled regions.}
\label{Fig:1}
\end{figure}

We consider SPDC in a planar waveguide in a $z$-cut, $x$-propagating potassium titanyl phosphate (KTP) substrate. In order to simultaneously realize two type-II SPDC processes we assume that the substrate has two periodical poling at two different slant angles \cite{7,8} as shown in Fig.~\ref{Fig:1}.  With appropriate poling periods it is possible to achieve simultaneously the following two down conversion processes from a horizontally polarized pump photon into a pair of signal and idler photons with the same frequencies $\omega_s$ and $\omega_i$:
\begin{align}\nonumber
1: & \qquad H_{p0} \rightarrow H_{s0} + V_{i0} \nonumber \\
2: & \qquad H_{p0} \rightarrow V_{s0} + H_{i0}. \nonumber
\end{align}

\begin{figure}[t]
\centering
\includegraphics[width=6cm]{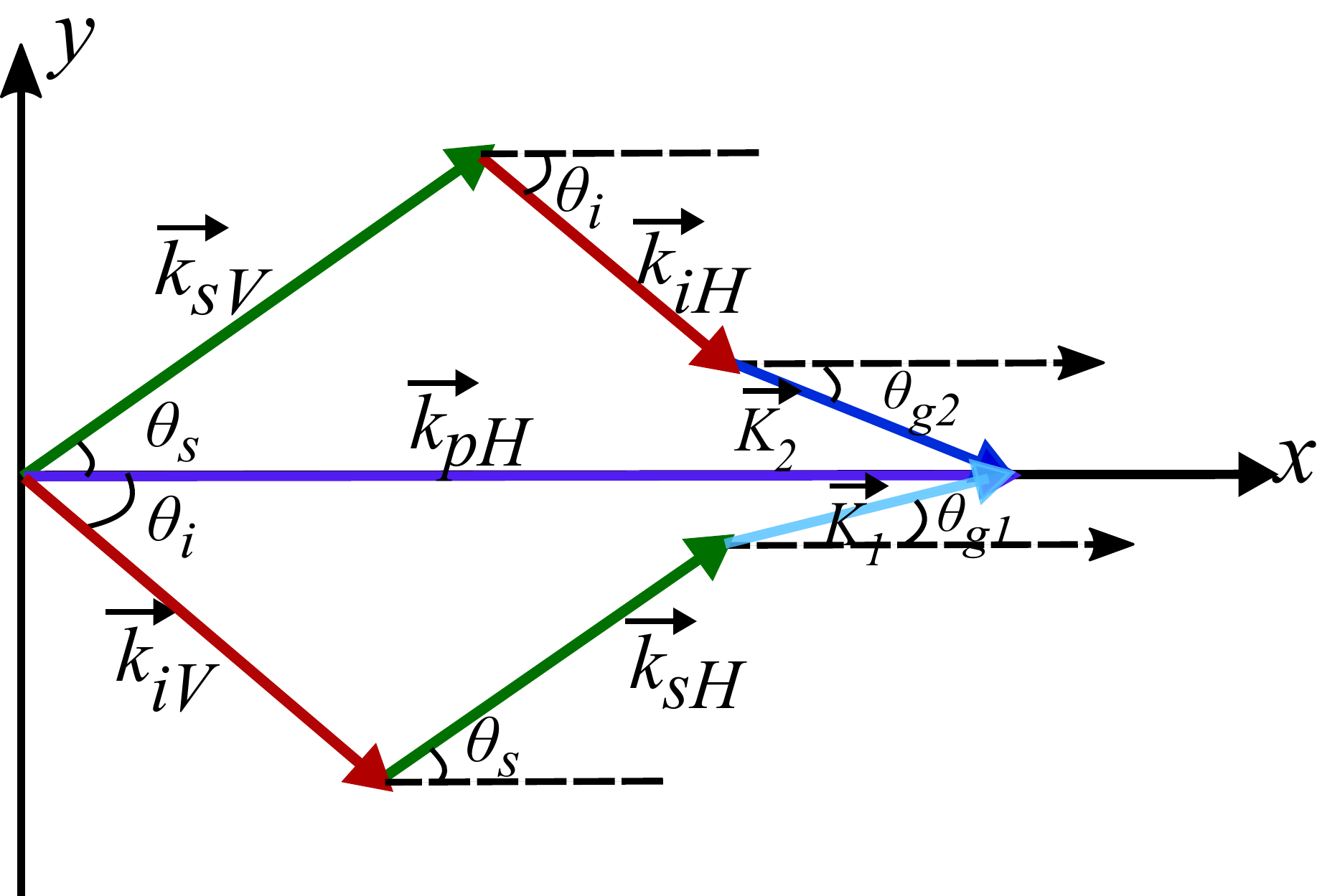}
\caption{Wave vector diagram for the two SPDC processes.}
\label{Fig:2}
\end{figure}

We will refer to $H_{p0(s0,i0)}$ and $V_{p0(s0,i0)}$ as horizontal and vertical polarization states of pump (signal, idler) in the fundamental $(0)$ mode respectively, with dominant components of the electric field oriented respectively along the plane of the waveguide and perpendicular to the plane of the waveguide. The vector diagrams showing the phase matching conditions for the two SPDC processes is shown in Fig.~\ref{Fig:2}. The figure shows the possibility of achieving pairs of horizontal and vertical polarized signal and idler pairs along the two chosen pairs of directions.
In such a case the output is expected to be a polarization entangled state given by (see Sec.~\ref{Sec:ANALYSIS}):
\begin{equation}
\ket{\Psi}=\eta\int d\omega_s\big[{f_{HV}}\ket{H_{s0},V_{i0}} + {f_{VH}}\ket{V_{s0},H_{i0}} \big].
\label{Eq:1}
\end{equation}
where $f_{HV}$ and $f_{VH}$ are coefficients defined in Eqs.~(\ref{Eq:10b}).

The phase mismatches of the two QPM conditions corresponding to the two processes are given as:
\begin{subequations}
\begin{align}
\Delta\vec{k}_1=\vec{k}_{pH}-\vec{k}_{sH}-\vec{k}_{iV}-\vec{K}_1\\
\Delta\vec{k}_2=\vec{k}_{pH}-\vec{k}_{sV}-\vec{k}_{iH}-\vec{K}_2 
\end{align}
Here,
\begin{align}
\vec{k}_{pH}=\beta_{pH}\hat{x}; \nonumber \\
\vec{k}_{sH(V)}=\beta_{sH(V)}(\cos\theta_s\hat{x}+\sin\theta_s\hat{y});  \nonumber \\
\vec{k}_{iH(V)}=\beta_{iH(V)}(\cos\theta_i\hat{x}+\sin\theta_i\hat{y}); \nonumber \\
\vec{K}_j=K_j(\cos\theta_{gj}\hat{x}+\sin\theta_{gj}\hat{y})
\label{Eq:2c}
\end{align}
\end{subequations}
where it is assumed that the signal and idler pairs corresponding to the two orthogonal polarizations appear along the same angle as shown in Fig.~\ref{Fig:2}. As shown in Fig.~\ref{Fig:2}, $\vec{k}_{pH}$ is the horizontally polarized pump wave vector along the $x$ - axis; $\vec{k}_{sH(V)}$ and $\vec{k}_{iH(V)}$ are the wave vectors corresponding to the horizontally (vertically) polarized signal and idler modes making angles $\theta_s$ (emission angle of signal) and $\theta_i$ (emission angle of idler) respectively with $x$ -axis; $\overrightarrow{K_1}$ and $\vec{K_2}$ are the grating vectors making angles $\theta_{g1}$  and $\theta_{g2}$  with $x$- axis respectively. In Eq.~(\ref{Eq:2c}), $\beta_{\alpha{m}}={\textstyle2{\pi}}\frac{n_{\alpha{m}}}{\lambda_\alpha}$ is the propagation constant, where $\alpha=p,s,i$ for pump, signal and idler respectively; $m= H, V$ for the horizontal and vertical polarization respectively; $\lambda_{p(s,i)}$ is the pump (signal, idler) wavelength and $n_{\alpha{m}}$ is the effective indices at different frequencies and polarization.

The phase mismatches corresponding to the $x$ and $y$-components for the two processes are given by: 
\begin{subequations}
\begin{flalign}
\Delta{k_x^{(HV)}}&=\beta_{pH}-\beta_{sH}\cos\theta_s-\beta_{iV}\cos\theta_i -K_1\cos\theta_{g1}  
\label{Eq:3a}\\
\Delta{k_y^{(HV)}}&=\beta_{sH}\sin\theta_s+\beta_{iV}\sin\theta_i -K_1\sin\theta_{g1} 
\label{Eq:3b}\\
\Delta{k_x^{(VH)}}&=\beta_{pH}-\beta_{sV}\cos\theta_s-\beta_{iH}\cos\theta_i -K_2\cos\theta_{g2}
\label{Eq:3c} \\
\Delta{k_y^{(VH)}}&=\beta_{sV}\sin\theta_s+\beta_{iH}\sin\theta_i -K_2\sin\theta_{g2} 
\label{Eq:3d}
\end{flalign}
\end{subequations}

In order that the output state defined by Eq.~(\ref{Eq:1}) is a polarization entangled state, the signal and idler pairs in the two processes have to appear at the same pair of angles i.e. direction of emission of horizontally and vertically polarized signal (or idler) must be same (see Fig.~\ref{Fig:1}). With a proper design of waveguide geometry and QPM slant gratings the coefficients $f_{HV}$ and $f_{VH}$ can be made equal, thus providing the possibility of producing maximally entangled state with maximum efficiency in a planar waveguide. 

We will show in Sec.~\ref{Sec:NUMERICAL SIMULATIONS} that down conversion process implemented in planar waveguide has higher efficiency and enhanced pair rate in comparison to bulk nonlinear crystal. We will also show that by an appropriate choice of the angle made by the grating vectors, it is possible to satisfy the quasi phase matching conditions corresponding to both components and thus achieve high efficiency SPDC into planar waveguide modes of the waveguide. In addition, we will show that by changing the pump wavelength, it is possible to generate different frequency pairs of entangled photons which will exit at different angles $(\theta_s,\theta_i )$ thus providing us the tunability of the SPDC process.
\section{ANALYSIS}
\label{Sec:ANALYSIS}
In this section we will provide a description of SPDC process in a periodically poled single moded planar waveguide at signal and idler wavelength for generation of polarization entangled photon pairs. 

We consider a pump to have a Gaussian transverse profile of beam waist $W_p$ along the $y$-direction and traveling along the $x$-direction in a dual periodically poled planar waveguide with the optic axis along $z$- axis(Fig.~\ref{Fig:1}) . We assume the pump to be described by a classical field as it is assumed to be strong. Thus the electric field at pump is given by:
\begin{subequations}
\begin{align}
\vec{E}_{pH}=\frac{1}{2}\big[\xi_{pH}(\vec{r},t)+\xi^*_{pH}(\vec{r},t)\big]\\
\xi_{pH}(\vec{r},t)=A_{p0}u_p(x,y,z)e^{i(k_{pH}x-\omega_pt)}
\label{Eq:4b}
\end{align}
\end{subequations}
Here, $u_p(x,y,z)$ represents the spatial distribution of pump described as:
\begin{align*}
u_{p}(x,y,z)=\left({\frac{1}{\sqrt\mu}}\right)\exp{\left(-\frac{y^2}{\mu{W_p^2}}\right)}\Psi_{p}^{(H)}(z) 
\end{align*}
where, $\mu= 1+i\frac{2x}{k_{pH}W_p^2}$\\
and,  $A_{p0}$ is the amplitude of pump given as:
\begin{align*}
 A_{p0}=\left({\frac{2}{\pi}}\right)^{\frac{1}{4}}\sqrt{\frac{2P_p}{W_pc\epsilon_0n_{pH}}}
\end{align*}
where, $ P_p$ is the pump power, $\epsilon_0$ is the free space permittivity and $c$ is the speed of light in vacuum.  

The quantized electric fields at signal and idler corresponding to different polarization are represented by the following equations:
\begin{subequations}
\begin{flalign}
\hat{E}_{s(i)m}=\frac{1}{2}&\big[\hat{\xi}_{s(i)m}(\vec{r},t)+\hat{\xi}^\dagger_{s(i)m}(\vec{r},t)\big]\\
\hat{\xi}_{s(i)m}(\vec{r},t)=i&\sum_{m=H,V}\sum_l\sum_{\vec{k}_{s(i)m}^{(l)}}\sqrt{\frac{2\hbar\omega_{s(i)}}{({n^{(l)}_{s(i)m})}^2\epsilon_0A}}\Psi_{s(i)l}^{(m)}(z)\nonumber\\
&\times e^{i(\vec{k}_{s(i)m}^{(l)}.\vec{r}_t-\omega_{s,i}t)}\hat{a}_{s(i)m}(\vec{k}_{s(i)m}^{(l)})
\label{Eq:5b}
\end{flalign}

Since we are considering the single moded planar waveguide at signal and idler wavelength, thus, $l=0$ and hence by putting $\vec{k}_{s(i)m}^{(0)}=\vec{k}_{s(i)m}$, $n^{(0)}_{s(i)m}=n_{s(i)m}$ and $\hat{a}_{s(i)m}(\vec{k}^{(0)}_{s(i)m})=\hat{a}_{s(i)m}(\vec{k}_{s(i)m})$, we get
\begin{flalign}
\hat{\xi}_{s(i)}(\vec{r},t)=i&\sum_{m=H,V}\sum_{\vec{k}_{s(i)m}}\sqrt{\frac{2\hbar\omega_{s(i)}}{n^2_{s(i)m}\epsilon_0A}}\Psi_{s(i)0}^{(m)}(z)\nonumber\\&\times e^{i(\vec{k}_{s(i)m}^{(l)}.\vec{r}_t-\omega_{s,i}t)}\hat{a}_{s(i)m}(\vec{k}_{s(i)m})
\label{Eq:5c}
\end{flalign}
\end{subequations}
Here, $\vec{k}_{s(i)m}=(k_{s(i)x}^{(m)},k_{s(i)y}^{(m)})$, $\omega_{p,s,i}$ are pump, signal and idler frequencies, $A=L_xL_y$; $L_x$ and $L_y$ are the length of the quantization volume along the $x$ and $y$ directions respectively, $\hat{a}_{s(i)m}$ and $\hat{a}^\dagger_{s(i)m}$  represent the annihilation and creation operators of the generated signal (idler) photons of transverse wave vectors $\vec{k}_{s(i)m}$ corresponding to fundamental spatial mode and the $m=H, V$ polarization and $\psi_{p(s,i)l}^{(m)}(z)$ is the normalized modal field profile for pump (signal, idler) along the $z$– direction of the $H$- and $V$-polarized $l$-modes such that $\int |\psi_{p(s,i)l}^{(m)}|^2dz=1$. We have abbreviated $\vec{r}=(x,y,z)$ and $\vec{r_t}=(x,y)$ for denoting the three and two dimensional position vectors respectively.

The interaction Hamiltonian is given by \cite{9,10,11}  
\begin{subequations}
\begin{eqnarray}
\hat{H}_{int}(t)=\frac{-\epsilon_0}{2}\iiint d(x,y)\big(\xi_{pH}\hat{\xi}^\dagger_{sm}\hat{\xi}^\dagger_{in} +h.c.\big)dxdydz \nonumber \\
\label{Eq:6a}
\end{eqnarray}
where, $h.c.$ represents Hermitian conjugate, $d$ is the nonlinear coefficient and $L$ is the length of crystal along the propagation axis $\hat{x}$. Here, spatial dependence of $d$ is included to take into account the peridoic poling of crystal. Substituting Eqs.~(\ref{Eq:4b}) and (\ref{Eq:5c}) in Eq.~(\ref{Eq:6a}), we get the following expression for the interaction Hamiltonian:
\begin{multline}
\hat{H}_{int}(t)=\frac{\sqrt{\pi}W_{p}LA_{p0}d_{eff}\hbar\sqrt{\omega_s\omega_i}}{A} \sum_{\substack{m,n=H,V\\m\neq n}}\sum_{\vec{k}_{sm},\vec{k}_{in}}\left[\frac{I_z^{(mn)}}{n_{sm}n_{in}}\right.\\  \left.\times\phi_{mn}(\vec{k}_{sm},\vec{k}_{in})e^{i(\omega_p-\omega_s-\omega_i)t}\hat{a}_{sm}^\dagger(\vec{k}_{sm})\hat{a}_{in}^\dagger(\vec{k}_{in})+h.c.\right]
\label{Eq:6b}
\end{multline}
\end{subequations}
Here, $d_{eff}$ is the effective nonlinear coefficent and $\phi_{mn}(\vec{k}_{sm},\vec{k}_{in})$ is the phase matching function given as:
\begin{subequations}
\begin{flalign}
\phi_{mn}(\vec{k}_{sm},\vec{k}_{in})=&\exp\left(-\frac{(\Delta{k}_y^{(mn)}W_p)^2}{4}\right)\nonumber\\
&\times\sinc\left(\frac{\Delta{k}^{(mn)}L}{2}\right)e^{i\frac{\Delta{k}^{(mn)}L}{2}}
\label{Eq:7a}
\end{flalign}
where, 
\begin{flalign*}
\Delta{k}_y^{(mn)}&=k_{sy}^{(m)}  +k_{iy}^{(n)}-K_j\sin\theta_{gj}\\
\Delta{k}^{(mn)}&=\Delta{k}_x^{(mn)}-\frac{(\Delta{k}_y^{(mn)})^2}{2k_{pH}}\\
&=k_{pH}-k_{sx}^{(m)}-k_{ix}^{(n)}-K_j\cos\theta_{gj}-\frac{(\Delta{k}_y^{(mn)})^2}{2k_{pH}}\\
k_{s(i)x}^m&=\beta_{s(i)m}\cos\theta_{s(i)} ;  k_{s(i)y}^m=\beta_{s(i)m}\sin\theta_{s(i)}
\end{flalign*}
$I_z^{(mn)}$ is the overlap integral between the fundamental modes of pump, signal and idler for $m, n =H,V$ and $m \neq n$, and is described by:
\begin{flalign}
I_z^{(mn)}=\int \Psi_{p0}^{(H)}(z)(\Psi_{s0}^{(m)}(z))^*(\Psi_{i0}^{(n)}(z))^*dz
\label{Eq:7b}
\end{flalign}
\end{subequations}

Now, in accordance with the interaction picture, the overall output state is given as:   
\begin{flalign}
\ket{\Psi}&=\ket{0_s,0_i}+\frac{1}{i\hbar}\int dt \hat{H}_{int}(t)\ket{0_s,0_i} \nonumber\\
&=\ket{0_s,0_i}+\ket{\Psi_1}
\label{Eq:8}
\end{flalign}                                                    
The $\ket{0_s,0_i}$ state correspond to vacuum state i.e. no signal and idler photon.\\ 
Now, replacing $\sum_{(\vec{k}_{sm},\vec{k}_{in})} \rightarrow \frac{A^2}{(2\pi)^4}\int d\vec{k}_{sm}\int d\vec{k}_{in}$ and $d\vec{k}_{sm}=\frac{n_{s(i)m}}{c}d\omega_sdk_{sy}^{(m)}$ and assuming pump to be monochromatic with a single frequency, $\omega_p=\omega_s+\omega_i$, we get the overall normalized two photon entangled state (neglecting the vacuum state) as:
\begin{subequations}
\begin{flalign}
\ket{\Psi_1}&=\frac{\eta}{i}\int d\omega_s[C_{HV}(\omega_s)\hat{a}_{sH}^\dagger\hat{a}_{iV}^\dagger+C_{VH}(\omega_s)\hat{a}_{sV}^\dagger\hat{a}_{iH}^\dagger]\ket{0_s,0_i}\nonumber\\
&=\frac{\eta}{i}\int d\omega_s\big[{C_{HV}}\ket{H_{s0},V_{i0}} + {C_{VH}}\ket{V_{s0},H_{i0}} \big].
\label{Eq:9a}
\end{flalign}  
where, 
\begin{flalign}
C_{mn}(\omega_s)&=\sqrt{\omega_s(\omega_p-\omega_s)}I_z^{(mn)}\iint\phi_{mn}(\vec{k}_{sm},\vec{k}_{in}) dk_{sy}^{(m)}dk_{iy}^{(n)} 
\label{Eq:9b}\\
\eta&=\frac{Ad_{eff}LW_pA_{p0}}{2\sqrt\pi(2\pi c)^2}
\label{Eq:9c}
\end{flalign}
\end{subequations}

The wave function given by Eq.~(\ref{Eq:9a}) describes the general two - photon SPDC state i.e. entire field emerging from planar waveguide into free space.  However, in order to obtain the entangled photon state along the two chosen pair of angles $\theta_s$ and $\theta_i$ in the x- y plane, the output state is coupled into channel waveguide and thus, the coefficient $C_{mn}$ is integrated within a range of $\theta_{s(i)} \pm \delta\theta$ ; where, $\delta\theta$ is determined by width of the channel waveguide positioned at an angle $\theta_{s(i)}$. Thus the two photon entangled state along an angle is given as:
\begin{subequations}
\begin{flalign}
\ket{\Psi_2}=\frac{\eta}{i}\int d\omega_s\big[{f_{HV}}\ket{H_{s0},V_{i0}} + {f_{VH}}\ket{V_{s0},H_{i0}} \big].
\label{Eq:10a}
\end{flalign}
Here,
\begin{flalign}
f_{mn}(\omega_s)=&\sqrt{\omega_s(\omega_p-\omega_s)}I_z^{(mn)}\beta_{sm}\beta_{in}\nonumber\\ &\times \int _{\theta_s-\delta\theta} ^{\theta_s +\delta\theta}\int _{\theta_i-\delta\theta} ^{\theta_i+\delta\theta}\phi_{mn} d\theta_sd\theta_i
\label{Eq:10b}
\end{flalign}
\end{subequations}

The concurrence E, an entanglement monotone, can be calculated as \cite{12,13,14}:
\begin{equation}
E=\frac{2|{\int d\omega_s}\left[f_{HV}(\omega_s)f_{VH}^*(\omega_s)\right]|}{\int d\omega_s[|f_{HV}(\omega_s)|^2+|f_{VH}(\omega_s)|^2]]}
\label{Eq:11}
\end{equation}

The output state will be maximally (polarization) entangled state for $E =1$ i.e.$|f_{HV}(\omega_s)|=|f_{VH}(\omega_s)|$; $\forall$  $\omega_s $ or using a narrow - band wavelength filter to limit the range of $\omega_s$ and will be generated with maximum efficiency for $\Delta{k}_y^{(mn)}=0$ and $\Delta{k}_x^{(mn)}=0; m,n=H,V$ and $m \neq n$.

Next, we derive an expression for the power of signal generated in planar waveguide through type - II SPDC process and for this, we will follow the same approach as mentioned in Ref. \cite{10,15}. We first calculate the transition rate using the Fermi’s Golden Rule. To calculate the transition rate, we need to find the density of states. 

The number of $H (V)$ - polarized signal and $V (H)$ - polarized idler states in the element $d^2\vec{k}_{sH(V)}d^2\vec{k}_{iV(H)}$ is given as:
 \begin{subequations}
\begin{flalign}
dN_{1(2)}=\frac{A^2}{(2\pi)^4}d^2\vec{k}_{sH(V)}d^2\vec{k}_{iV(H)}
\label{Eq:12a}
\end{flalign}
Therefore, the density of states in planar waveguide is given as:
\begin{flalign}
\rho^{(P)}_{1(2)}=\frac{A^2n_{sH(V)}n_{iV(H)}}{(2\pi)^4{\hbar}c^2}dk_{sH(V)y}dk_{iV(H)y}d\omega_s
\label{Eq:12b}
\end{flalign}
Thus, the transition rate for the $ j= 1$ and $2$ processes is:
\begin{flalign}
T_{1(2)}=\frac{2\pi}{\hbar}|\bra{H(V)_s,V(H)_i}\hat{H}_{int}\ket{0_s,0_i}|^2\rho^{(P)}_{1(2)}
\label{Eq:12c}
\end{flalign}
\end{subequations}
The $\ket{H(V)_s,V(H)_i} = \hat{a}_{sH(V)}^\dagger\hat{a}_{iV(H)}^\dagger\ket{0_s,0_i}$ state corresponds to final state having one signal and one idler photon corresponding to the fundamental spatial modes in horizontal (vertical) and vertical (horizontal) polarization respectively.

The down converted signal power in a frequency interval, $d\omega_s$ or wavelength interval,  $d\lambda_s$ is, $dP_s=\hbar\omega_sT$. Thus, we get the following expression for signal power for $m,n=H,V$ and $m \neq n$: 
\begin{flalign}
dP_{s}^{(mn)}=2(2\pi)^{3/2}&\frac{{\hbar}cd^2P_pL^2W_p\left(I_z^{(mn)}\right)^2}{\epsilon_0n_{pH}n_{sm}n_{in}\lambda_s^4\lambda_i}\nonumber\\
&\times\iint dk_{iy}^{(n)}dk_{sy}^{(m)}|\phi_{mn}|^2 d \lambda_s
\label{Eq:13}
\end{flalign}
The radiated signal power in bulk, $dP_s^{(B)}$ is given as \cite{9}: 
\begin{flalign}
dP_{s1(2)}^{(B)}=\frac{2{\hbar}cd^2P_pL^2W_p^2I_{B1(2)}}{\epsilon_0n_{pH}n_{sH(V)}n_{iV(H)}\lambda_s^4\lambda_i}d\lambda_s
\label{Eq:14}
\end{flalign}
where,
\begin{flalign*}
I_{B1(2)}=\iiiint &\exp\left(-\frac{(\Delta{K^2_{y1(2)}}+\Delta{K^2_{z1(2)}})W_p^2}{2}\right)\\
&\sinc^2\left(\frac{\Delta{K_{x1(2)}}L}{2}\right)dk_{iy}dk_{iz}dk_{sy}dk_{sz}
\end{flalign*}
$\Delta{K_{xj}}, \Delta{K_{yj}}$ and $\Delta{K_{zj}}$ is the phase mismatch along the $x, y$ and $z$ directions respectively for $j=1$ and $2$  processes in bulk .

In order to make coefficients $f_{HV}$ and $f_{VH}$ to be equal, periodic poling of substrate is needed. For generation of polarization entangled photon pairs through type-II collinear $(<1^{\circ})$ and non-degenerate SPDC process we require two linear QPM gratings to satisfy the two longitudinal phase matching conditions only but for non collinear emission $(>1^{\circ})$ of polarization entangled photon pairs, either two slanted periodical poling of substrate or one slanted periodical poling of substrate in addition to one linear QPM grating is needed to satisfy all the four QPM conditions simultaneously. Also, the transverse phase matching condition dictates that if the emission angle of signal is in first quadrant then idler emission angles must be in fourth quadrant and vice versa. It may also be worth mentioning here that if the two slant grating are used then they must be aligned in two opposite directions with respect to propagation axis $\hat{x}$ in order to satisfy the two transverse phase matching conditions simultaneously for non collinear emission.

In planar waveguide, an angular degree of freedom is available with respect to the phase matching conditions of the two processes (see Eqs.~(\ref{Eq:3a}) - (\ref{Eq:3d})), so that tuning of the pump wavelength leads to the generation of maximally entangled photon pairs having specific combinations of signal and idler wavelengths emitted along a particular pair of emission angles. In contrast, channel waveguides in nonlinear materials can confine light in both transverse  directions, implying a restriction to only one translation degree of freedom, i.e. the propagation direction of the interacting photons, so no angular degree of freedom is available and thus tunability of generated photon pairs is not possible. 

In addition, in the case of planar waveguides signal and idler photons are physically separated and so there is no need of any device to separate them unlike in the case of channel waveguide.
\section{NUMERICAL SIMULATIONS}
\label{Sec:NUMERICAL SIMULATIONS}
In order to validate our analysis, we present numerical results for an ion exchanged potassium titanyl phosphate (KTP) \cite{16,17} planar waveguide of depth, $d = 2.0 ~\mathrm{\mu m}$ with a step index refractive profile. For the numerical simulations, the values of the KTP substrate refractive indices $(n_s)$ for different wavelengths and different polarizations were calculated using Sellmeier equations given in \cite{18} and the refractive index difference $(\Delta{n})$ for a waveguide is taken to be 0.02 \cite{16}.

\begin{figure}[t]
\centering
\includegraphics[width=6cm]{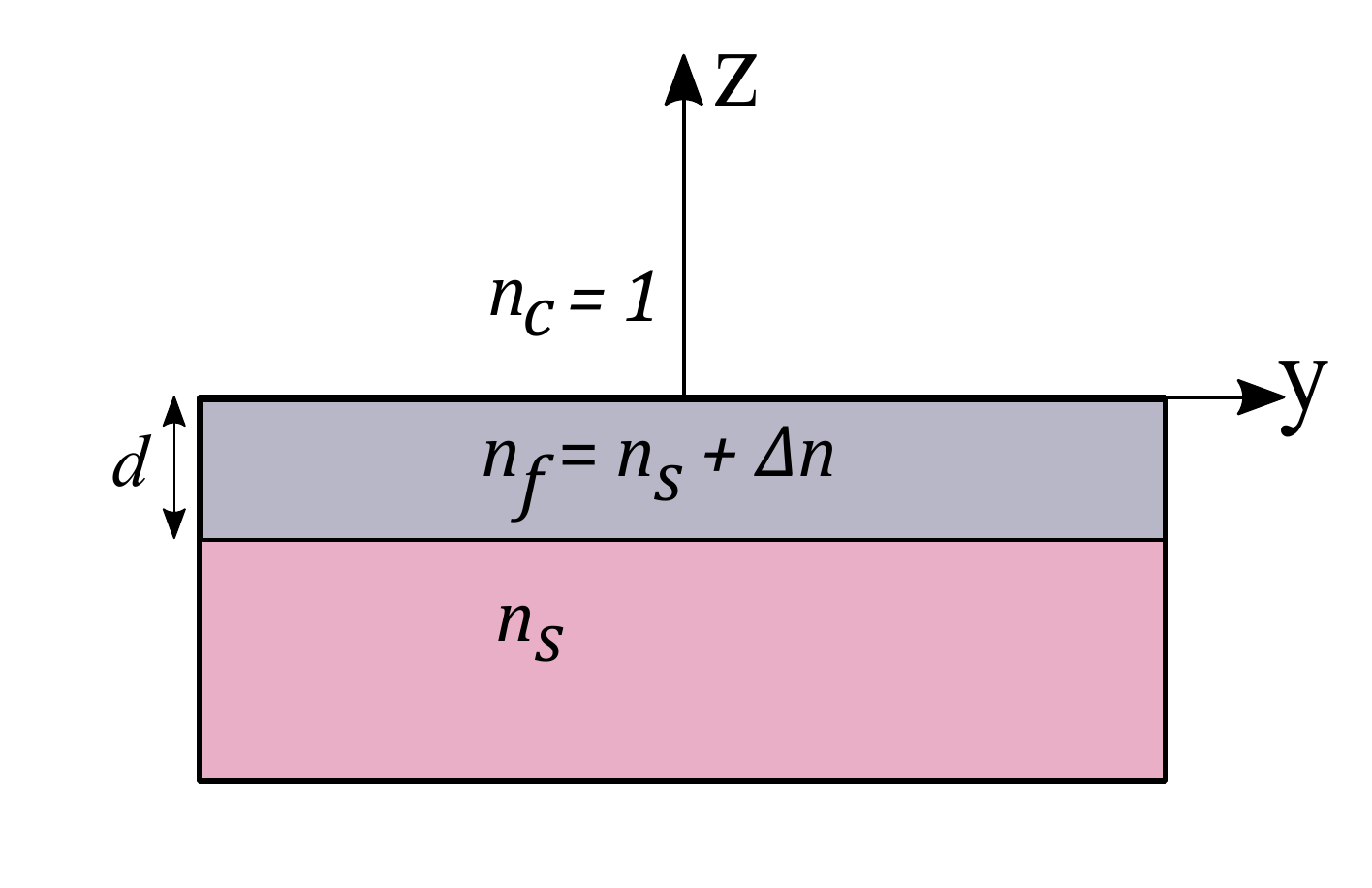}
\caption{Cross sectional view of the planar waveguide }
\label{Fig:3}
\end{figure}

We have carried out the modal analysis \cite{19} for planar waveguide of depth, $d$ (Fig.~\ref{Fig:3}) to calculate the eigenmodes and thus, obtained the propagation constant of $H$-polarized and $V$-polarized fundamental mode at pump, signal and idler wavelength by solving eigenvalue equation of TE mode and TM mode respectively.

Using the modal field distributions of the $H$- and $V$-polarized fundamental mode of pump, signal and idler so evaluated from modal analysis of planar waveguide, we have calculated the overlap integrals defined by Eq.~(\ref{Eq:7b}) and it is found that the overlap integrals for both the processes are almost equal. 

\begin{figure}[t]
\centering
\includegraphics[width=8cm]{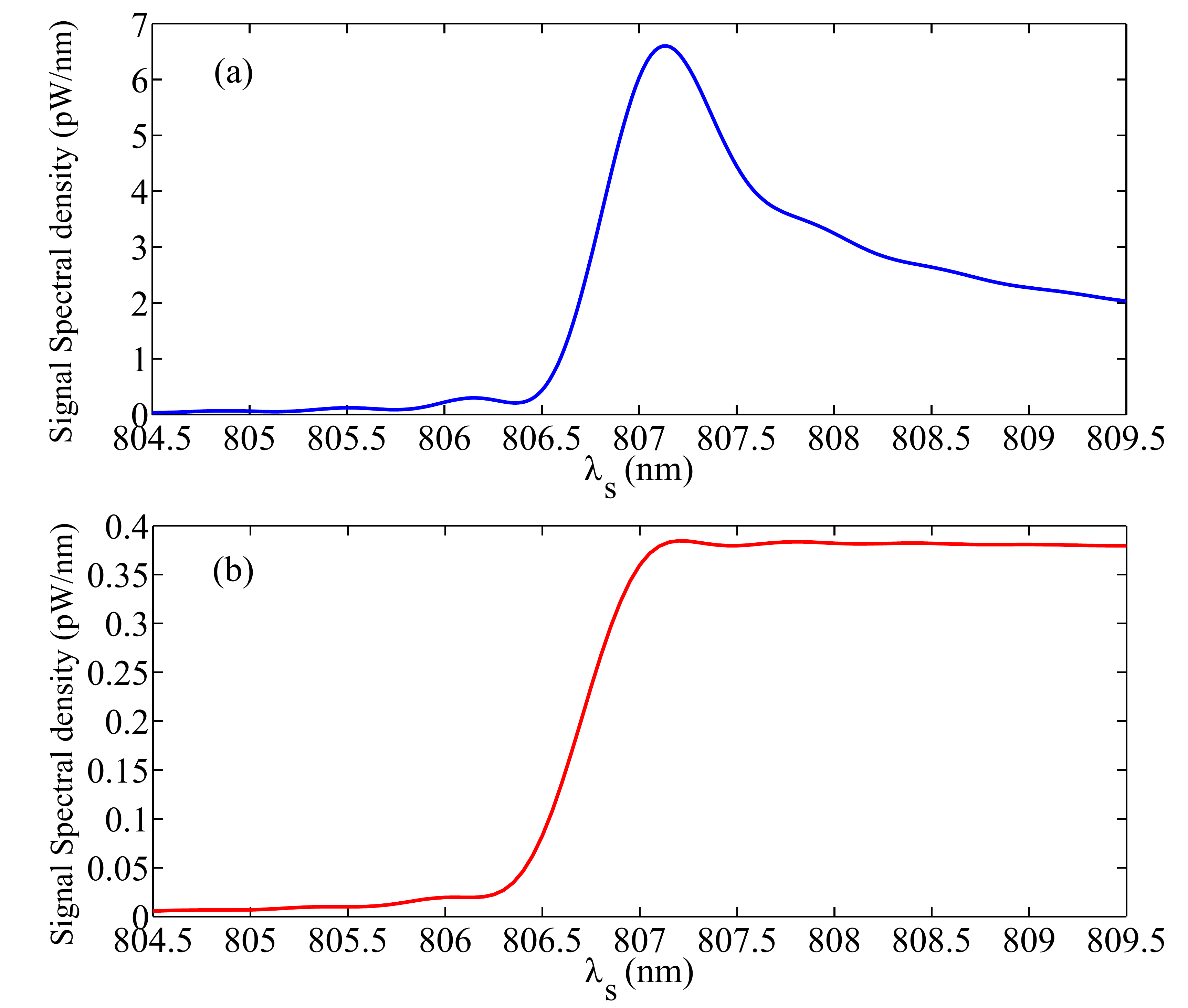}
\caption{Variation of signal power density with wavelength in (a) planar waveguide (b) bulk  }
\label{Fig:4}
\end{figure}

We first demonstrate that the production rate of the down – converted photon pairs generated through SPDC process in a periodically poled KTP (PPKTP) planar waveguide is more than in periodically poled bulk KTP nonlinear crystal. For this, we consider a type-II collinear and non-degenerate SPDC process assuming a 405 nm pump beam of $W_p =100~\mathrm{\mu m}$ beam waist with a pump power of 1 mW in a PPKTP of length, $L$ = 10 mm. The variation of signal power density with signal wavelength is shown in Fig.~\ref{Fig:4}(a) for periodically poled planar waveguide of grating period, $\Lambda = 9.0 ~\mathrm{\mu m}$. Fig.~\ref{Fig:4}(b) shows the power density in case of periodically poled bulk substrate with QPM grating period,  $\Lambda = 10.06 ~\mathrm{\mu m}$  (this period is different from the waveguide case as in this case the effective index is just the bulk index of the substrate).

It can be seen from Figs.~\ref{Fig:4}(a) and \ref{Fig:4}(b) that signal power density in the planar waveguide is larger by a factor of about 10 compared to that in bulk which in turn leads to higher rate of generation of SPDC pairs in planar waveguide in comparison to bulk.

\begin{figure}[t]
\centering
\includegraphics[width=8cm]{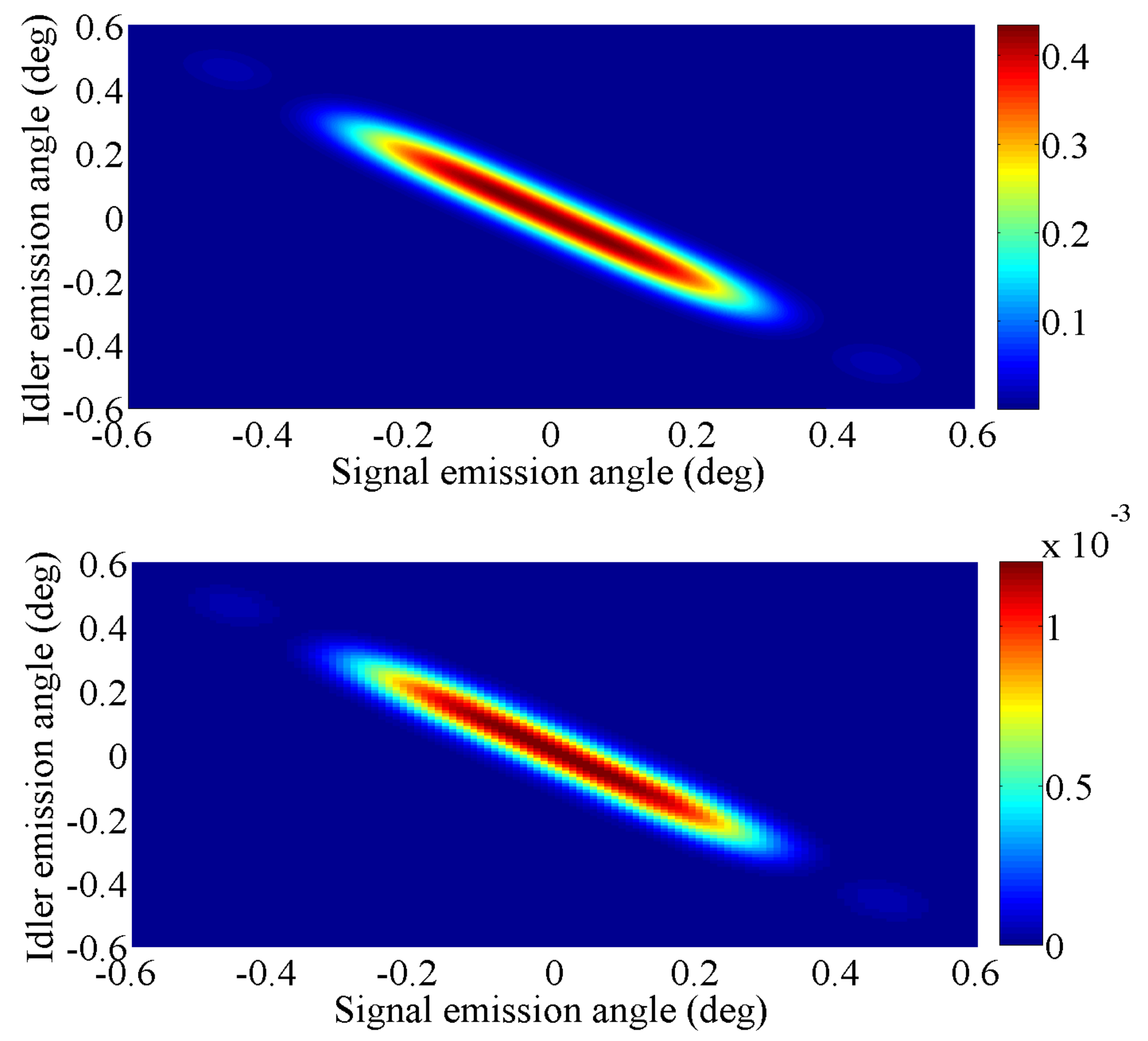}
\caption{Variation of signal power density with emission angles in (a) planar waveguide (b) bulk. In case of bulk cyrstal, signal power density is calculated at a zero polar angle $(\phi = 0)$. Notice the highly different scales for efficiencies in the two cases.}
\label{Fig:5}
\end{figure}

Next, we compare the signal power density vs emission angles for both planar waveguide and bulk crystals for a collinear phase matching at a pump and  signal wavelength of 405 nm and 807 nm respectively. From Figs.~\ref{Fig:5}(a) and \ref{Fig:5}(b) , we can see that the signal power density for planar waveguide is much larger ($\sim$300 times) than that of bulk.

Now, we show the generation of polarization entangled photon pairs through type II non – collinear and non-degenerate SPDC process in KTP having one linear grating and one slant periodic domain reversal grating. We calculate the photon flux in an angular region of $\pm0.034^\circ$ (corresponding to $6~\mathrm{\mu m}$ wide channel waveguide) along an appropriate pairs of signal and idler emission angles. 

\begin{figure}[t]
\centering
\includegraphics[width=8cm]{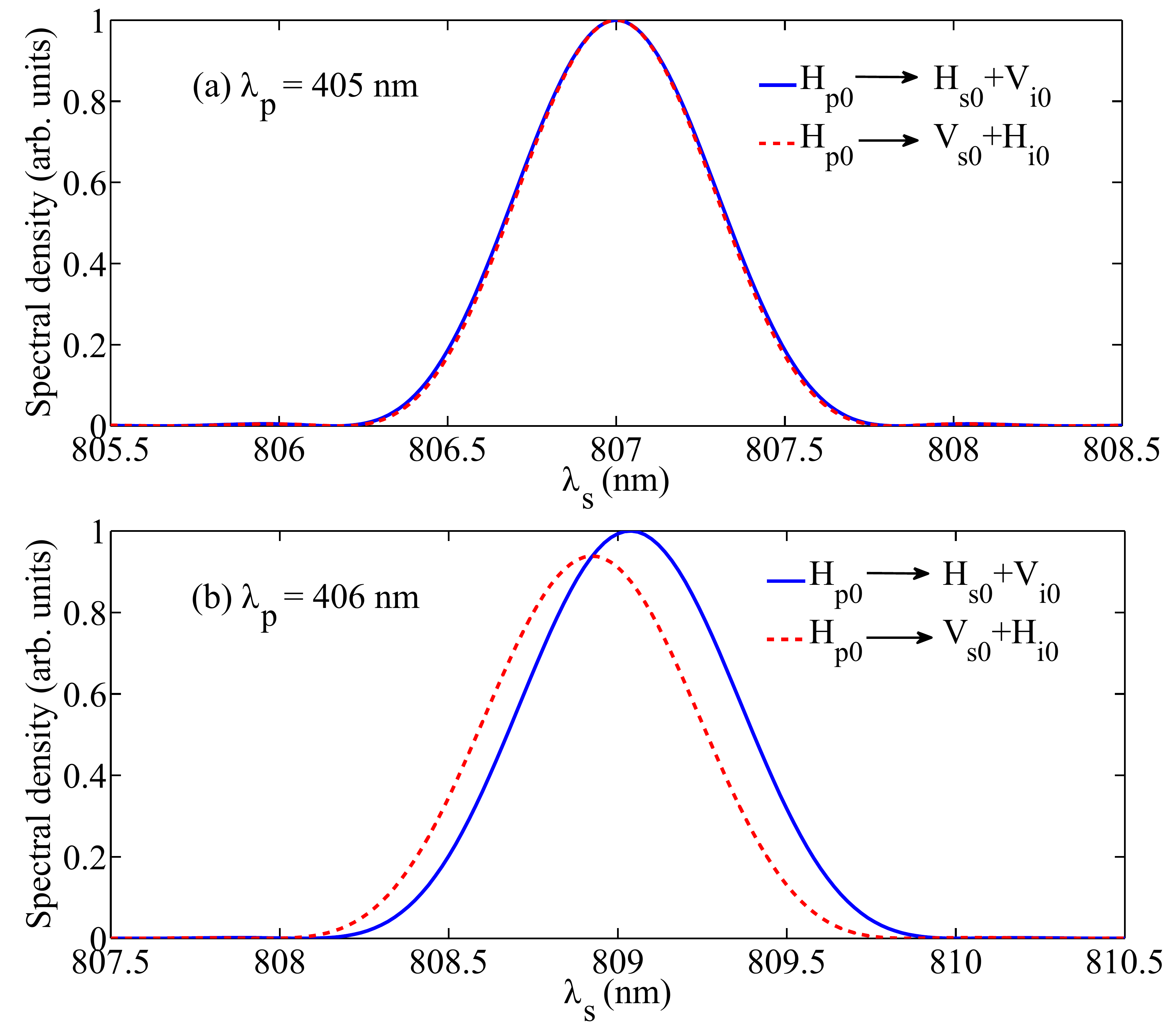}
\caption{Plot of output spectra for two processes as a function of signal wavelength for a pump wavelength (a) 405~nm (b) 406~nm }
\label{Fig:6}
\end{figure}

Simulations have been carried out for generation of maximally entangled photon pairs  for a pump wavelength of 405 nm, pump beam waist, $W_p =100~\mathrm{\mu m}$ , signal emission angle, $\theta_s=2.5^\circ$  and waveguide parameters as:  waveguide length = 10 mm; depth =$2.0 ~\mathrm{\mu m}$. We optimize the signal wavelength and idler emission angle for slant grating angle, $\theta_{g1}=0^\circ$ which dictates the conservation of transverse phase matching function of process 1.  Thus the conservation of transverse QPM condition of process 1 for signal wavelength, $\lambda_{s} = 807$~nm corresponds to $\theta_i=-(\theta_s-0.1)$. Now, we chose the QPM grating period of first grating = $8.62~\mathrm{\mu m}$, period of second grating = $8.75~\mathrm{\mu m}$ and slant grating angle, $\theta_{g2} = -4.64^\circ$   so that the QPM conditions for both the processes can be simultaneously satisfied and the polarization entangled photon pairs appear with maximum probability along the pair of angles  ($\theta_s, \theta_i$) = ($2.5^\circ, -2.4^\circ$) corresponding to the pair of wavelengths ($\lambda_{s},\lambda_{i}$)  = (807~ nm, 813.02 ~nm)(see Fig.~\ref{Fig:6}(a)). It can be seen from Fig.~\ref{Fig:6}(a) that the two processes have overlapping spectra with identical bandwidths of 1.3~ nm. Thus, $|f_{HV}(\omega_s)|=|f_{VH}(\omega_s)|$ over an entire region of spectral overlap and this leads to maximally entangled state with a concurrence, E=1. 

In context to the tunability of generated photon pairs in planar waveguide, we study generation of entangled photon pairs at different wavelengths along different pairs of angles using different pump wavelengths but with same gratings.Figure~\ref{Fig:6}(b) shows the normalized output spectra at the signal wavelength corresponding to both of the SPDC processes for pump wavelength of 406 nm. It can be seen that by  changing the pump wavelength to 406 nm entangled photon pairs can be generated with maximum efficiency for the wavelength pair of  ($\lambda_{s},\lambda_{i}$) =  (808.92~nm, 815.01~nm) at a different angle pair of ($\theta_s, \theta_i$) = ($2.94^\circ, -2.84^\circ$). Thus by using a narrow band wavelength filter at the signal wavelength of 808.92 nm, we can obtain  entangled photon pairs at the output with E$\approx$0.96 for a filter bandwidth of 0.1 nm. It is worth noting here that even by changing the pump wavelength by 1 nm  maximally polarization entangled state can still be generated along the  different pair of emission angles.

It may be worth mentioning here that the entangled photons pairs can also be generated for higher emission angles with larger tunability of generated photon pairs by changing the grating periods but at a cost of efficiency of generated photon pairs.

\begin{figure}[t]
\centering
\includegraphics[width=8cm]{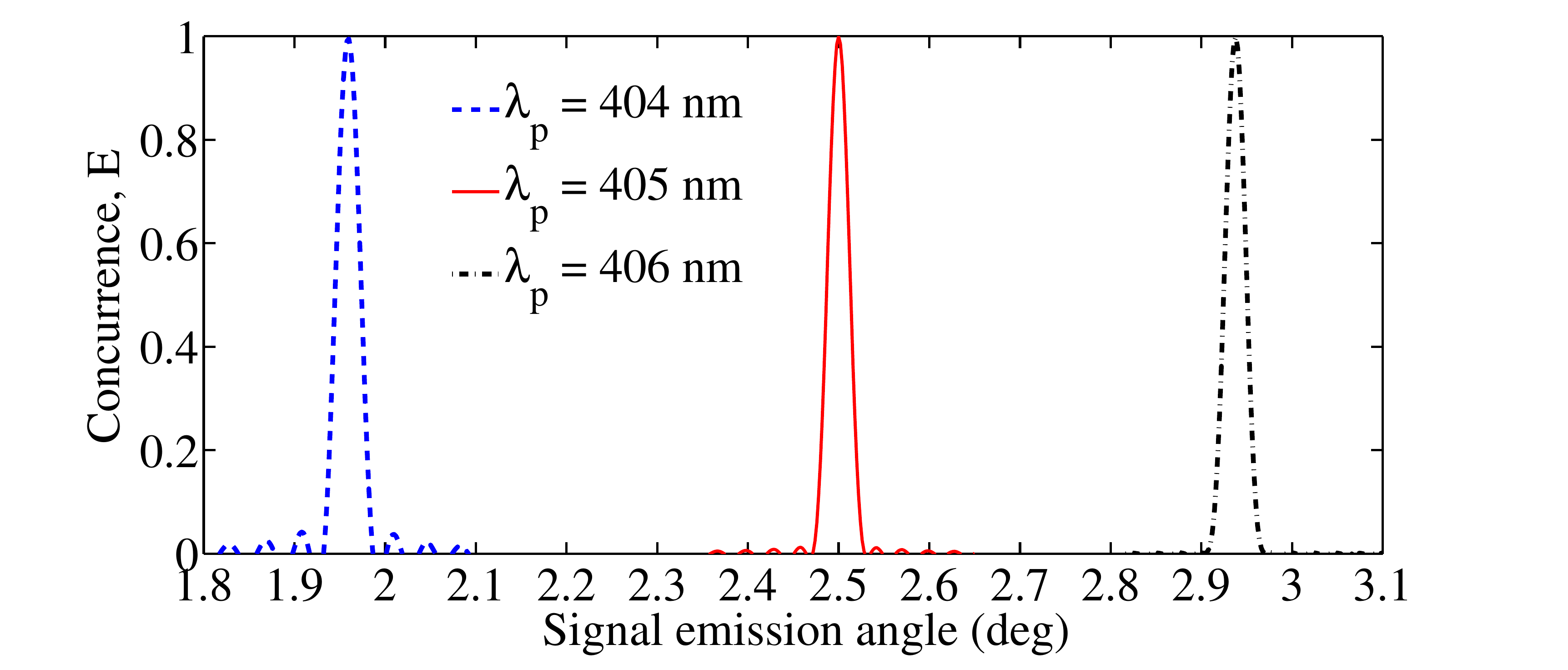}
\caption{Variation of concurrence, E as function of signal emission angles}
\label{Fig:7}
\end{figure}

Now, in order to demonstrate the feature of tunability we plot the concurrence as a function of signal emission angles with idler emission angles, $\theta_i=-(\theta_s-0.1)$ for three different pump wavelengths(see Fig.~\ref{Fig:7}). 
It can be clearly seen from  Fig.~\ref{Fig:7} that the photons are maximally entangled for different pump wavelength at different emissions angles exhibiting the tunability of the design.  
 
It is worth to emphasize here that in the case of channel waveguides, in view of the phase matching condition any change in pump wavelength by as much as 0.1~nm leads to reduction in SPDC process and does not lead to generation of maximally entangled photon pairs.  

\begin{figure}[t]
\centering
\includegraphics[width=8cm]{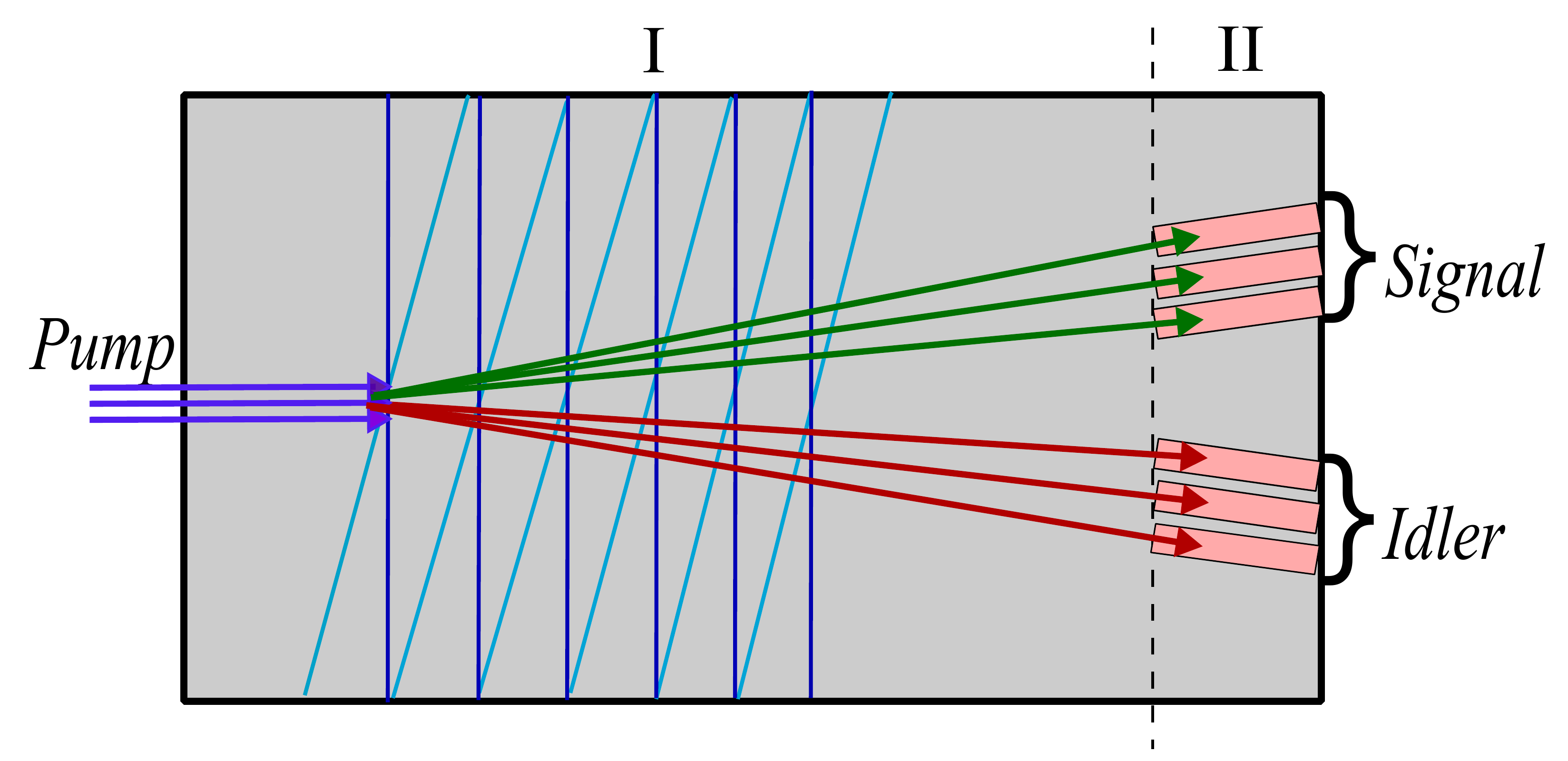}
\caption{Top view of the waveguide design for generating and collecting of polarization entangled photon pairs. Here, region (I) is a planar waveguide with two slant QPM gratings for generating different polarization entangled stats by changing the pump wavelength states and region (II) shows the collection of photon pairs along the different channel waveguide}
\label{Fig:8}
\end{figure}

Figure~\ref{Fig:8} shows a possible implementation of the proposal in integrated optic form. At the end of the planar waveguide section channel waveguides can be positioned at appropriate angles as per the design to which the generated photon pairs can get coupled (much like in the case of arrayed waveguide gratings used in optical fiber communication\cite{20}). The output of the channel waveguides can be coupled to optical fibers for further processing. Depending on the pump wavelength different wavelength pairs of entangled signal and idler wavelengths will exit from different pairs of channel waveguides.

In addition to this, planar waveguide configuration with appropriate QPM gratings can also lead to generation of hyper – entangled state which are simultaneously entangled in polarization as well as in path degree of freedom. Such a state is defined as:
\begin{equation}
\ket{\psi_3}=\big[\ket{s_I,i_{II}}+\ket{s_{II},i_I}\big]\otimes\big[\ket{H_{s0},V_{i0}}+\ket{V_{s0},H_{i0}}\big]
\label{Eq:15}
\end{equation}
Here, $s$ and $i$ correspond to signal and idler and the subscripts I and II corresponds to the path I and path II making an angle $\theta_s$ and $\theta_i$ with the pump beam direction respectively.

\begin{figure}[t]
\centering
\includegraphics[width=8cm]{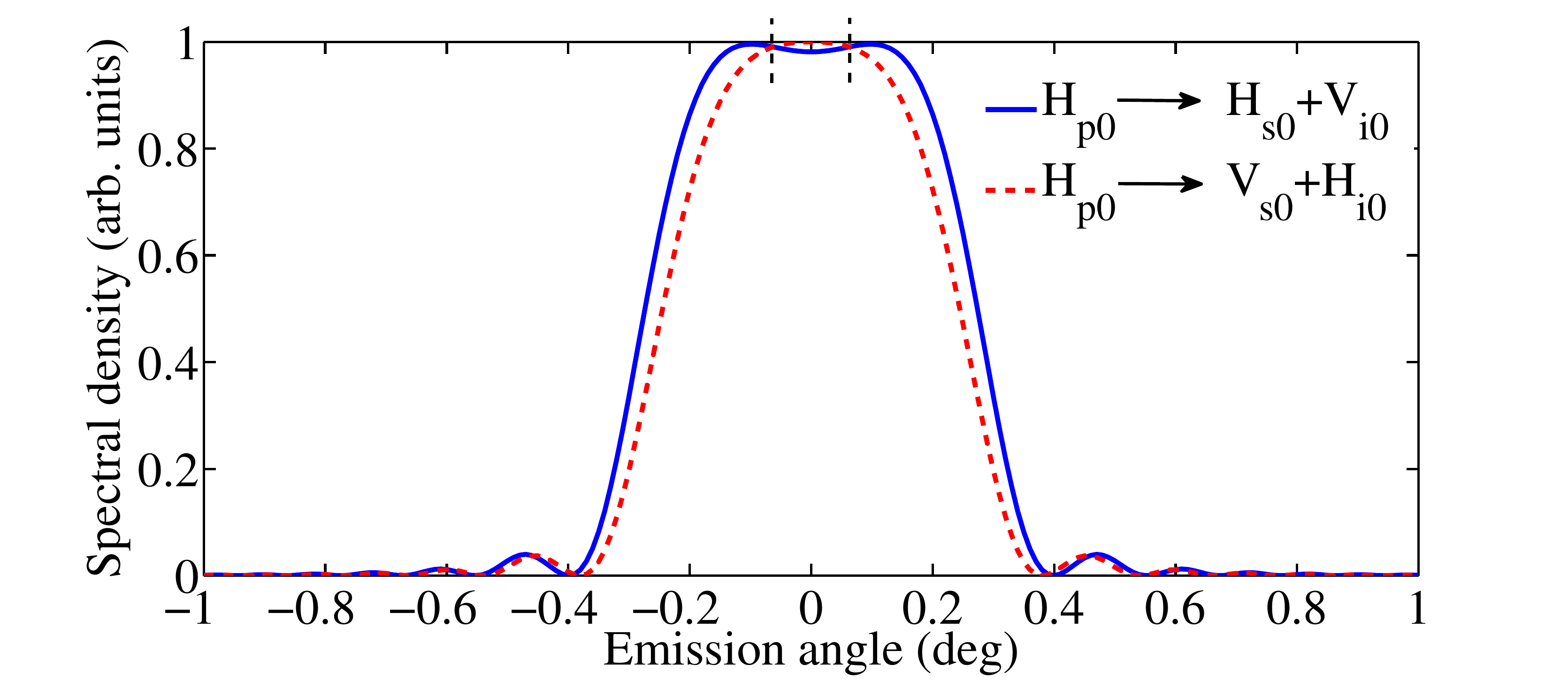}
\caption{Output spectra as a function of emission angles }
\label{Fig:9}
\end{figure}

As an example, we demonstrate here the generation of hyper - entangled photon pairs at 807~nm and 813.02~nm from a pump beam of wavelength 405~nm excited in a fundamental modes in a planar waveguide of depth $2.0~\mathrm{\mu m}$. For gratings of  periods $9.0~\mathrm{\mu m}$ and  $9.08~\mathrm{\mu m}$ having periodical poling parallel to propagation  direction, we show the variation of the normalized output spectra as a function of emission angle. It can be inferred from Fig.~\ref{Fig:9} that polarization entangled photon pairs can be simultaneously generated over a range of angles of $- 0.4^\circ$ to $ 0.4^\circ$ and can be generated with maximum efficiency at the pairs of emission angles $0.1^\circ$ and $-0.1^\circ$ corresponding to path I and path II respectively where the two curve intersect as shown by the dotted vertical lines in Fig.~\ref{Fig:9}. The state of the photon pairs generated in the pairs of paths is given by Eq.~(\ref{Eq:15}).

\section{CONCLUSION}
\label{Sec:CONCLUSION}
We have shown that planar waveguides provide us an optimized configuration for generation of polarization entangled photon pairs using SPDC with efficiencies higher than for bulk with the possibility of ease of collection like in channel waveguides with the additional feature of tunability. Such a design should find applications in integrated quantum optics. We have also shown that the planar waveguide configuration with appropriate QPM gratings can also lead to generation of hyper – entangled state unlike in case of channel waveguide.
\begin{acknowledgments}
The work of D.B. was supported by Ministry of Human Resource Department (MHRD), New Delhi, under the Senior Research Fellow (SRF) scheme.
\end{acknowledgments}


\end{document}